\documentclass[fleqn,usenatbib]{mnras}

\usepackage{newtxtext,newtxmath}

\usepackage[T1]{fontenc}
\usepackage{ae,aecompl}

\usepackage{graphicx}	
\usepackage{amsmath}	
\usepackage{amssymb}	

\newcommand{\kepler}{\textit{Kepler}}
\newcommand{\rosat}{\textit{ROSAT}}

\newcommand{\PLATO}{\textit{PLATO}}
\newcommand{\ngts}{{NGTS}}
\newcommand{\objname}{\mbox{NGTS\,J0308-2113}}
\newcommand{\longobjname}{\mbox{NGTS\,J030834.9-211322}}

\newcommand{\MADsymb}{$MAD$}

\newcommand{\teff}{$T_{\rm eff}$}

\newcommand{\largeenergy}{$5.4^{+0.8}_{-0.7}\times10^{34}$}
\newcommand{\smallenergy}{$2.6^{+0.4}_{-0.3}\times10^{34}$}

\title[G star superflares with \ngts]{Ground-based detection of G star superflares with \ngts}
\author[J. A. G. Jackman et al.]{
James A. G. Jackman,$^{1,2}$\thanks{E-mail: J.Jackman@warwick.ac.uk}
Peter J. Wheatley,$^{1,2}$\thanks{E-mail: P.J.Wheatley@warwick.ac.uk}
Chloe E. Pugh,$^{1,2}$\newauthor
Boris T. G\"ansicke,$^{1,2}$
Edward Gillen,$^{3}$
Anne-Marie Broomhall,$^{1,2,4}$
David J. Armstrong,$^{1,2}$\newauthor
Matthew R. Burleigh$,^{5}$
Alexander Chaushev,$^{5}$
Philipp Eigm\"uller,$^{6}$
Anders Erikson,$^{6}$\newauthor
Michael R. Goad,$^{5}$
Andrew Grange,$^{5}$
Maximilian~N.~G{\"u}nther,$^{3}$
James S. Jenkins,$^{7,8}$\newauthor
James McCormac,$^{1,2}$
Liam Raynard,$^{5}$
Andrew P. G. Thompson,$^{9}$
St\'{e}phane~Udry,$^{10}$\newauthor
Simon Walker,$^{1}$
Christopher A. Watson,$^{9}$
Richard G. West$^{1,2}$
\\
$^{1}$Dept. of Physics, University of Warwick, Gibbet Hill Road, Coventry CV4 7AL, UK\\
$^{2}$Centre for Exoplanets and Habitability, University of Warwick, Gibbet Hill Road, Coventry CV4 7AL, UK\\
$^{3}$Astrophysics Group, Cavendish Laboratory, J.J. Thomson Avenue, Cambridge CB3 0HE, UK\\
$^{4}$Institute of Advanced Studies, University of Warwick, Coventry CV4 7HS, UK\\
$^{5}$Department of Physics and Astronomy, Leicester Institute for Space and Earth Observation, University of Leicester, LE1 7RH, UK\\
$^{6}$Institute of Planetary Research, German Aerospace Center, Rutherfordstrasse 2, 12489 Berlin, Germany\\
$^{7}$Departamento de Astronomia, Universidad de Chile, Casilla 36-D, Santiago, Chile\\
$^{8}$ Centro de Astrof\'isica y Tecnolog\'ias Afines (CATA), Casilla 36-D, Santiago, Chile.\\
$^{9}$Astrophysics Research Centre, School of Mathematics and Physics, Queen's University Belfast, BT7 1NN, Belfast, UK\\
$^{10}$Observatoire Astronomique de l'Universit\'{e} de Gen\`{e}ve, 51 Ch. des Maillettes, 1290 Versoix, Switzerland\\
}

\date{Accepted XXX. Received YYY; in original form ZZZ}

\pubyear{2015}
\begin{document}
\label{firstpage}
\pagerange{\pageref{firstpage}--\pageref{lastpage}}
\maketitle
%
\begin{abstract}
We present high cadence detections of two superflares from a bright G8 star (V = 11.56) with the Next Generation Transit Survey (\ngts).
We improve upon previous superflare detections by resolving the flare rise and peak, allowing us to fit a solar flare inspired model without the need for arbitrary break points between rise and decay. Our data also enables us to identify substructure in the flares. From changing starspot modulation in the \ngts\ data we detect a stellar rotation period of 59 hours, along with evidence for differential rotation. We combine this rotation period with the observed \rosat\ X-ray flux to determine that the star's X-ray activity is saturated. We calculate the flare bolometric energies as \largeenergy and \smallenergy erg and compare our detections with G star superflares detected in the \kepler\ survey. We find our main flare to be one of the largest amplitude superflares detected from a bright G star. 
With energies more than 100 times greater than the Carrington event, our flare detections demonstrate the role that ground-based instruments such as \ngts\ can have in assessing the habitability of Earth-like exoplanets, particularly in the era of \PLATO.
\end{abstract}




\section{Introduction}
Stellar flares are explosive phenomena caused by reconnection events in a star's magnetic field \citep[e.g.][]{Benz10}. When previously observed from the ground, they have been synonymous with active M stars, which flare regularly and brightly compared to their quiescent flux. Yet it is well known that the Sun shows regular flaring behaviour, with flares being detected over a wide range of energies. These range from $10^{23}$ erg for ``nanoflares" \citep[][]{Parnell2000} up to approximately $10^{32}$ erg for the largest occurrences such as the Carrington event \citep[][]{Carrington_1859, Hodgson1859, Carrington_Energy}.
Observations of solar type stars, mainly with \kepler, have shown that much more energetic ``superflares'' of bolometric energies $10^{33}$ to $10^{36}$ erg are also possible \citep[e.g.][]{Shibayama13}. 

The discovery of Earth sized exoplanets in the habitable zones of their host stars \citep[e.g. TRAPPIST-1 and Proxima Centauri:][]{Gillon17,Anglada16} has given renewed importance to these superflares, in particular their effects on exoplanet habitability \citep[e.g.][]{Lingam17}. Previous studies have found that the increase in UV radiation associated with flares can result in ozone depletion \citep[][]{Segura2010}, changes to atmospheric composition \citep[][]{Venot16} and even biological damage \citep[e.g.][]{Estrela17}. These effects are relatively well studied for M dwarf hosts, however it is expected that in future \PLATO\ \citep[][]{PLATO14} will reveal habitable zone planets around K and G stars. Compared to flares from later-type counterparts, detections of superflares from G stars are relatively rare. To date, no G star superflares have been detected with a CCD detector from the ground, although several have been seen either visually, in photography or with vidicon detectors 
\citep[][]{Schaefer1989, Schaefer_2000}.

In recent years, observations with the \kepler\ satellite \citep[][]{Borucki2010} have captured greater numbers of superflare events from G-type stars. These have been from both the long (30 minutes) and short (1 minute) cadence modes. In the long cadence mode, \citet{Maehara2012} and \citet{Shibayama13} found 365 and 1547 superflares from 148 and 279 G-type stars respectively. In the 1-minute short cadence mode, 187 superflares from 23 solar-type stars were found by \citet{Maehara15}. From these detections, the statistical properties of superflares on G-type stars were considered, with \citet{Maehara2012} and \citet{Shibayama13} finding a power law distribution of occurrence rate against energy of superflares that is comparable to solar flares, and with  \cite{Maehara15} identifying a correlation between the e-folding flare duration (time from flare amplitude peak to 1/e of its initial value) and the bolometric flare energy.

\citet{Candelaresi14} also studied the occurrence rate of superflares from G dwarfs, as well as K and M dwarfs. They found the occurrence rate of superflares decreased with stellar effective temperature, and also peaked at a Rossby number of 0.1 (where Rossby number is the ratio of rotation period and convective turnover time). A Rossby number of 0.1 also corresponds to the rotation rate at which the X-ray emission of active stars saturates at 0.1 percent of the bolometric luminosity \citep[e.g][]{Pizzolato03,Wright11}. 

Previous studies have shown there may be a possible maximum limit on the energy that can be output by a G star superflare. \citet{Wu15} identified a saturation value of around $2 \times 10^{37}$ erg, using stars in the sample of \citet{Maehara2012} that displayed periodic modulation. 
Similar saturation behaviour was detected by \cite{Davenport16} from their sample of 4041 flaring stars, for example from the flaring G dwarf KIC 11551430. 
\citet{Davenport16} also found evidence for a weak correlation between flare luminosity and rotation period. 

While these detections have shown the statistical properties of these white light flares, their temporal morphology and its link to solar flare morphology has not been investigated. This is due to undersampling of the flare rise and peak from previous stellar flare surveys (mainly \kepler), particularly for shorter duration events. High cadence (< 1 minute) data are required 
in order to compare observed solar flares and stellar superflares.

In this paper we present the first ground based CCD detections of superflares from a G type star. These are some of the most well resolved superflares to date, with a higher cadence than all \kepler\ measurements and most ground based observations. We present our measurements of the stellar and flare parameters and make comparisons with previously detected G star flares. We also present our modelling of each flare using a solar inspired general flare model. 
\begin{table}
	\centering
	\begin{tabular}{|l|c|c|}
    \hline
    Property & Value & Reference\tabularnewline \hline
    $\textrm{RA}_{NGTS}$ & 03:08:34.9 \tabularnewline
    $\textrm{Dec}_{NGTS}$ & -21:13:22 \tabularnewline
    $\textrm{RA}_{NGTS}$ (Deg) & 47.14557 \tabularnewline
    $\textrm{Dec}_{NGTS}$ (Deg) & -21.22284 \tabularnewline
    $W4$ & 8.773 & 4\tabularnewline
	$W3$ & 9.699 & 4\tabularnewline
    $W2$ & 9.731 & 4\tabularnewline
    $W1$ & 9.699 & 4\tabularnewline
    $K_{s}$ & 9.768 & 1\tabularnewline
    $H$ & 9.865 & 1\tabularnewline
    $J$ & 10.216 & 1\tabularnewline 
    $i'$ & 11.174 & 3\tabularnewline 
    $r'$ & 11.356 & 3\tabularnewline
    $g'$ & 11.899 & 3\tabularnewline
    \textit{Gaia} $G$ & 11.354 & 2\tabularnewline
    $V$ & 11.562 & 3\tabularnewline
    $B$ & 12.291 & 3\tabularnewline
    $NUV$ & 16.943 & 5\tabularnewline
    $FUV$ & 20.666 & 5\tabularnewline
    ROSAT X-ray Count Rate (ct/s) & 0.042 & 6 \tabularnewline
   	$\mu_{RA}$ & $-1.2\pm1.1$ & 7 \tabularnewline
    $\mu_{DEC}$ & $-6.2\pm1.1$ & 7 \tabularnewline
	\hline
	\end{tabular}
    \caption{\label{tab:params} Properties of \objname. Coordinates are given in the J2000 system. References are as follows, 1. \citet{2MASS_2006}, 2. \citet{Gaia2016}, 3. \citet{APASS_14}, 4.\citet{ALLWISE2014}, 5. \citet{GALEX_05}, 6. \citet{Boller16} , 7. \citet{UCAC5_17} $NUV$, $FUV$, $i'$, $r'$, g' are AB magnitudes. Proper motions are in mas/yr.}
\end{table}

\begin{figure*}
	\includegraphics[width=\textwidth]{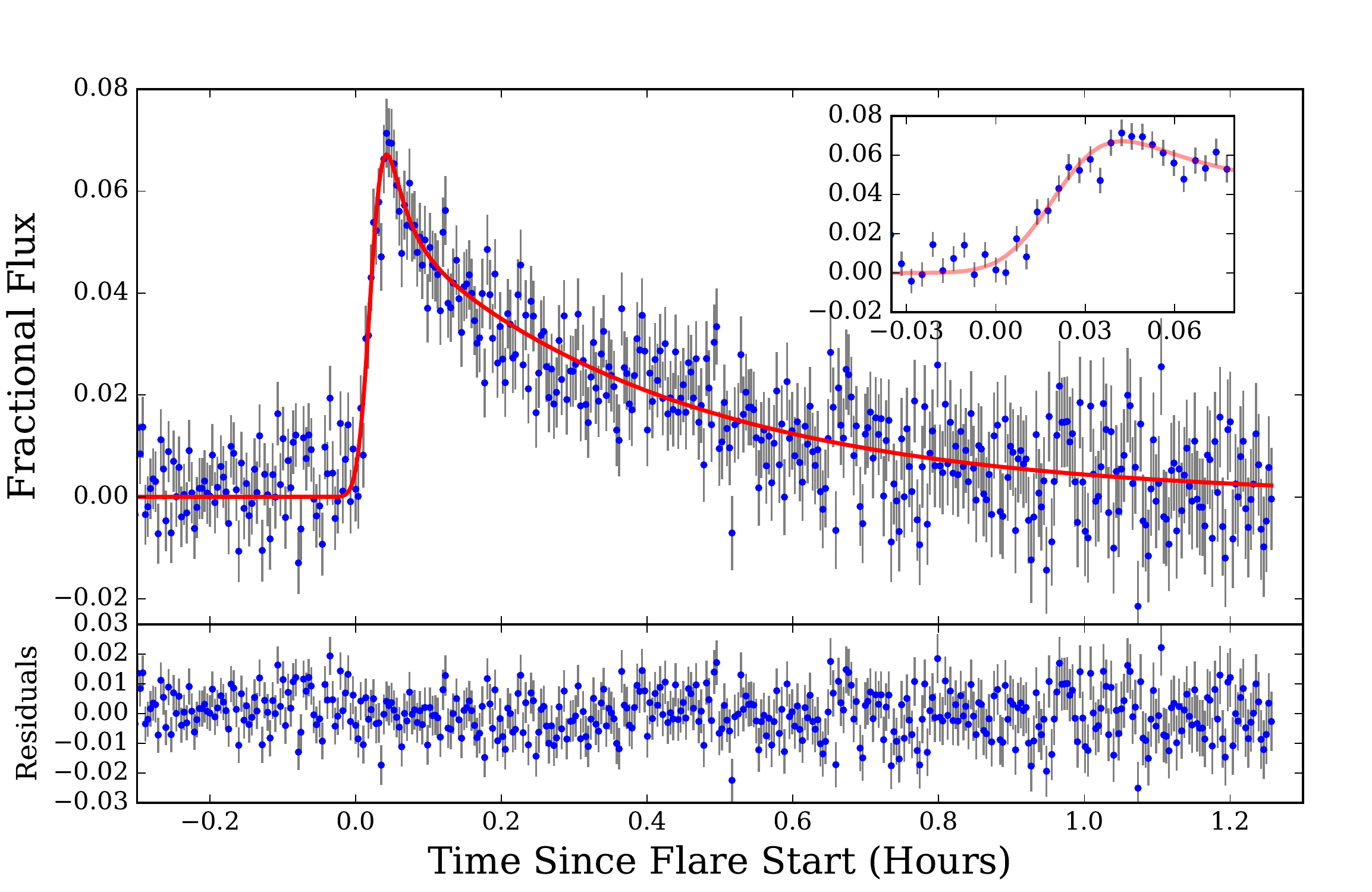}
    \caption{The superflare observed on \objname\, on 2016 January 3rd with best fitting model overlaid in red. The inset panel shows a zoomed in view of the data comprising the flare rise and peak, showing how the high cadence of \ngts\ has enabled us to resolve and fit to these regions. The bottom panel shows the residuals from our fitting. We note potential substructure in the flare decay around 0.7 hours and discuss this in Sect.\,\ref{sec:discussion}.}
    \label{fig:example_figure}
\end{figure*}

\section{Observations}
The data presented in this paper were collected with the Next Generation Transit Survey \citep[\ngts;][]{Wheatley18} over 80 nights between 2015 November 4th and 2016 February 25th. The two flares in this paper were detected on the nights of the 2015 December 17th and 2016 January 3rd. 
\ngts\ is a ground based transiting exoplanet survey, operating at Paranal. It has 12, 20\,cm f/2.8, optical telescopes, each with  a 520-890nm bandpass and exposure times of 10 seconds. Each single camera has a field of view of $\simeq$ 8\,deg$^2$. \ngts\ is designed to monitor bright (I$\leq$16) K and M stars in the search for exoplanet transits \citep[][]{Wheatley18}. Unlike \kepler, \ngts\ observes without a set target list, meaning that all stars in our field of view that are bright enough can be studied. 
Each \ngts\ field is observed intensively whenever visible, and 3--4 fields are observed per telescope each  year. 
With a total instantaneous field of view of 96\,deg$^2$, and 10\,s exposures, 
it is evident that \ngts\ 
is well suited
to measure flare statistical distributions along with temporal morphology.

\section{Data Analysis and Results}
\subsection{Flare search algorithm} \label{sec:detection}
When searching for flares, we started from the raw \ngts\ lightcurves and detrended them using a custom version of the {\scshape sysrem} algorithm \citep[][]{Sysrem_paper}. The full \ngts\ detrending is described by 
\citet{Wheatley18}.
For these detrended lightcurves, we applied an additional filter to remove frames which showed excess variance above an empirically defined limit, primarily to remove data adversely affected by clouds.
The timescale of stellar flares is minutes to hours \citep[e.g.][]{Poletto89},
so most
flares will have durations less than one night, and we searched for flares on a night-by-night basis.

In order to find flares in each night, we searched for 3 consecutive points greater than 6\,\MADsymb\ from the median of the night, where \MADsymb\ is the median absolute deviation. We have chosen \MADsymb\ as it is a robust measure of the variation within a night, and it is typically not strongly biased by the flare itself (a separate search is also carried out for flares that dominate the whole night). 
We applied no binning to the data, in order to fully utilise the time resolution of \ngts. Once the automated flagging procedure was complete, we inspected each flagged night visually and removed false positives. Examples of events which resulted in false positive flags include satellites passing through our aperture and high amplitude variable stars (e.g. RR Lyrae).

\subsection{Flare detection} 
Using the method from Sect.\,\ref{sec:detection} 
we detected a single flare, shown in Fig.\,\ref{fig:example_figure}, from the star \longobjname\  (\objname). This star has also previously been identified as 2MASS J03083496-2113222. After identifying this flare we visually inspected each night to search for lower amplitude flares which were not flagged. From this, we identified a second flare, shown in Fig.\,\ref{fig:secondary_flare}. 

To confirm the flares were not from a neighbouring source, we checked individual \ngts\ images from before and during the large flare, along with the positions of nearby stars from \textit{Gaia} and 2MASS. The nearest source identified is from \textit{Gaia}, a 20.658 magnitude star 10.5 arcseconds (2.1 pixels) away, placing it within our aperture. However, \ngts\ images reveal no shift in centroid position during the flare, and no light entering from outside the aperture, making us confident the flares are from \objname.

\subsection{Stellar Properties} \label{sec:stellar_prop}
\begin{figure}
	\includegraphics[width=\columnwidth]{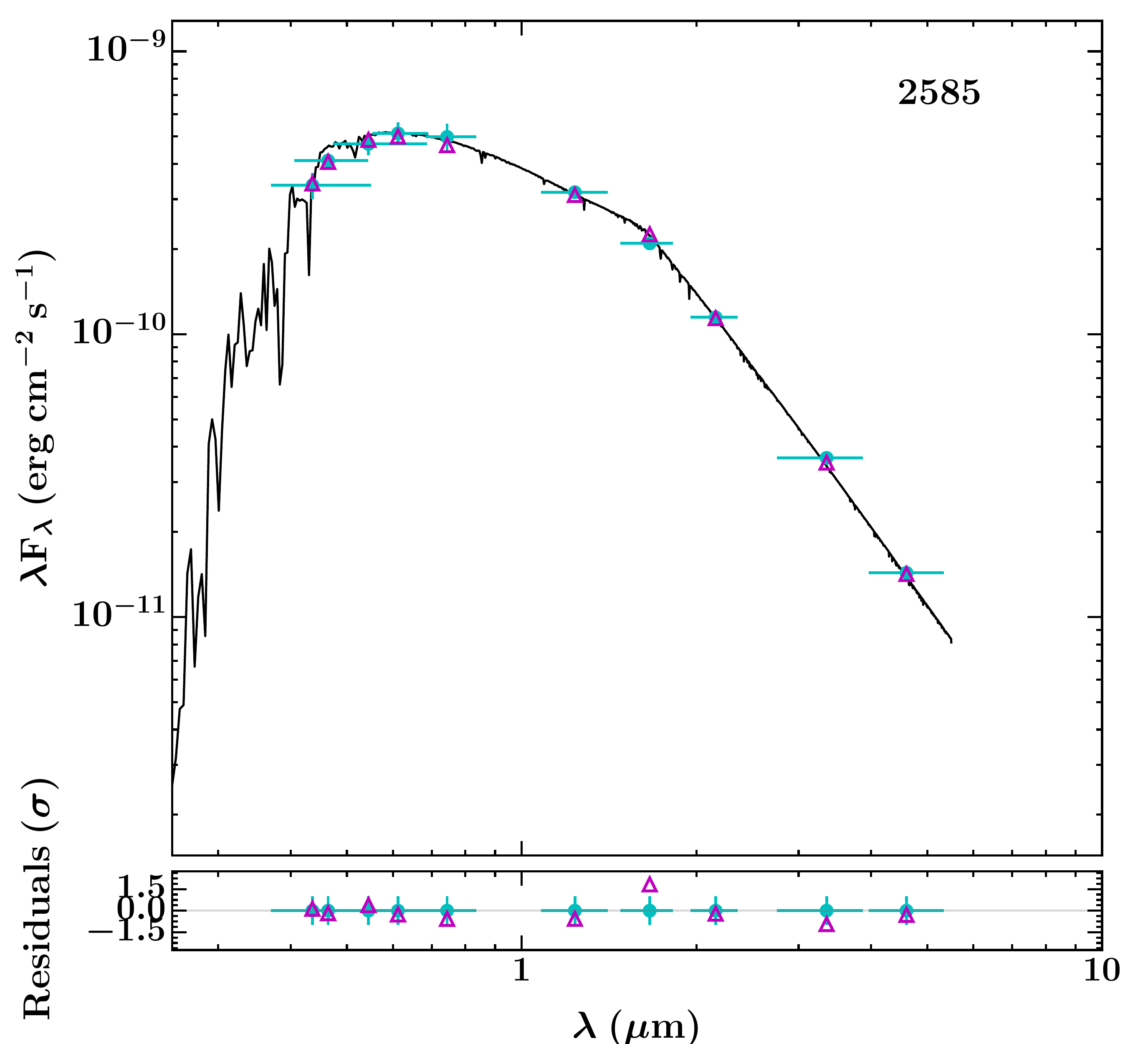}
    \caption{SED fit of \objname\ using the magnitudes listed in Tab.\,\ref{tab:params}, matching best to a G8 spectral type.}
    \label{fig:SED_fit}
\end{figure}

To determine the stellar parameters for \objname\ we performed SED fitting using the broadband photometry listed in Tab.\,\ref{tab:params}. These photometric values were obtained as part of the standard \ngts\ cross-matching pipeline \citep[][]{Wheatley18}. We use the SED modelling method described in \citet{Gillen17}, with the BT-SETTL and PHOENIX v2 model atmospheres. The SED fit is shown in Fig.\,\ref{fig:SED_fit}. We see no IR-excess that might indicate \objname\ is a pre-main sequence star.
From the SED fit we determine the effective temperature $T_{\rm eff} = 5458\,^{+108}_{-85}$ K. We then use the information presented in Table 5 of \citet{Pecaut13} to identify the spectral type as G8. As a check on the spectral type we can also use the stellar colours with Tables 3 and 4 of \citet{Covey07}, which confirm the G8 spectral type. 
To determine the
stellar radius, we assume the star is main sequence and use the empirical radius-temperature relation from equation 8 of \citet{Boyajian12,Boyajian17}, determined from mass-radius calculations for 33 stars of spectral type between G5V and M5.5V. 
We calculate our stellar radius as 0.81 $\pm$ 0.04 R$_{\sun}$. 
To estimate the uncertainty on the radius
we use the median absolute deviation of 0.031 R$_{\sun}$ from the \citet{Boyajian12} fit and combine it with the upper error for our stellar temperature.

To check this source was not a giant star we compared the reduced proper motion, $H_J$ against $J-H$ colour \citep[e.g.][]{Gould03}. Using the proper motion values from Tab.\,\ref{tab:params}, we calculate $H_J$= -0.78 and $J-H$= 0.35. We use the criteria for dwarf/giant classification from \citet{Collier07} to rule out the possibility this star is a giant.

We also note that this star was detected in X-rays with \rosat. The detection of X-rays from this source is a sign of an active stellar corona \citep[][]{Boller16}.

\subsection{Stellar Rotation} \label{sec:stellar_rotation}
\begin{figure}
	\includegraphics[width=\columnwidth]{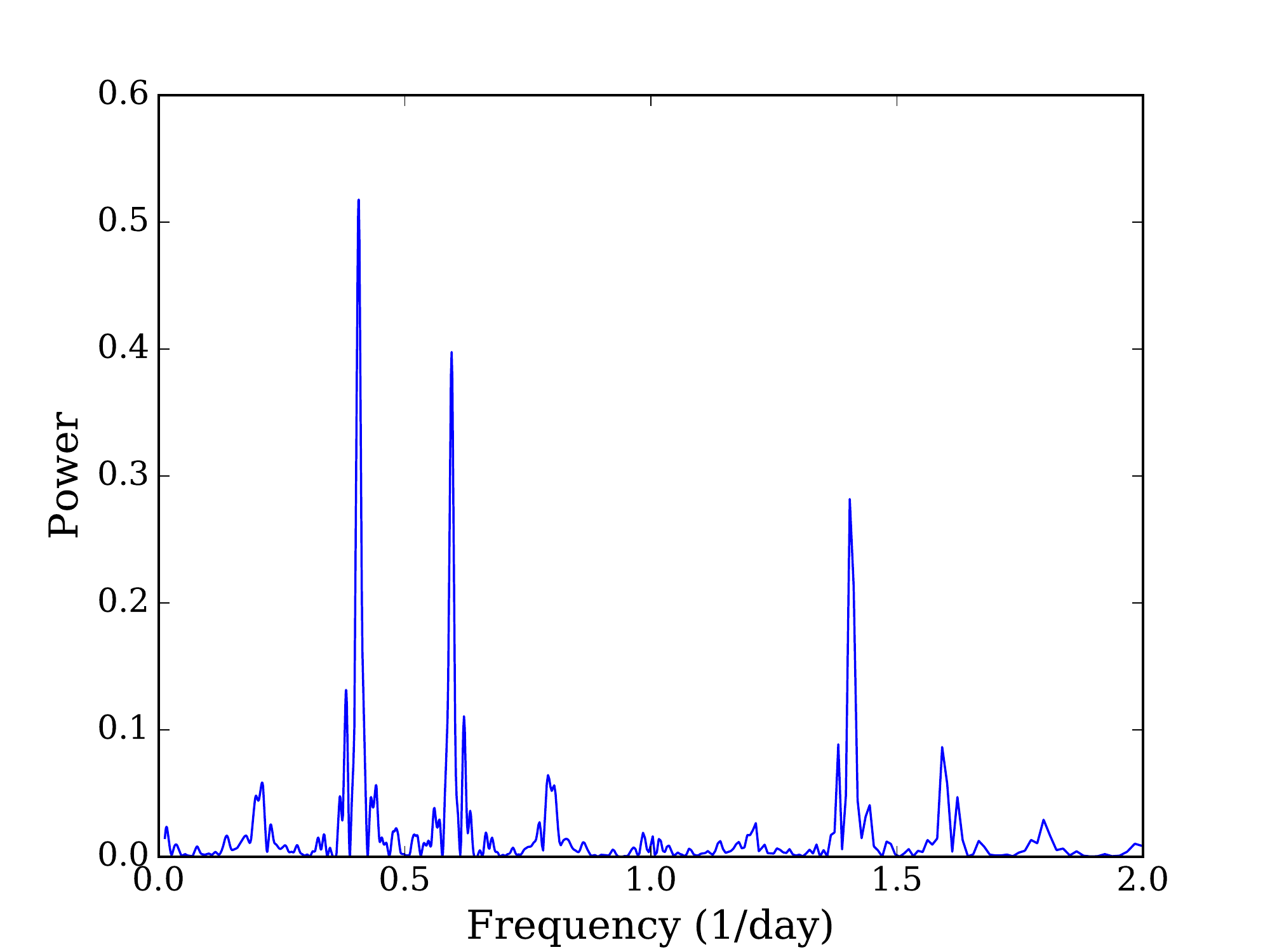}
    \caption{Lomb-Scargle periodogram for our full lightcurve. Here we show frequencies between 0 and 2 day$^{-1}$. Note the 1 day alias of the peak groups. The largest peak corresponds to our detected 59 hour period.}
    \label{fig:ls}
\end{figure}

The NGTS lightcurve of \objname\ shows periodic flux variations, which we attribute to starspots moving across the visible disc of the star. We use this behaviour to determine the rotation period of the star, using a Lomb-Scargle periodgram. To do this, we use the \textsc{astropy} package LombScargle \citep{Astropy13} and test for 20000 periods spaced between 13 seconds and 80 days. We mask the flares from our lightcurve when performing this analysis. Our periodogram for the whole time series of  \objname\ is shown in Fig.\,\ref{fig:ls}, from which 
the period of the main peak is 59.09$\pm$0.01 hours (0.41 $d^{-1}$). We calculate the uncertainty on this period by fitting a sine wave to the data. 
Using the analysis from \citet{Baluev08} we determine the false alarm probability (FAP) of this peak to be negligible, a result of the high amount of data.
We also note a second peak at 40 hours, which we found to be an alias by 
performing an identical Lomb Scargle analysis on a sine wave of period 59 hours with the same time sampling as our lightcurve. 

This short spin period implies that \objname\ must be a relatively young star, most likely less than $\sim$600 Myr old, through comparing to the observed spin-age relations of open clusters \citep[e.g.][]{Ardestani17,Douglas17,Stauffer16}.

\begin{figure}
	\includegraphics[width=\columnwidth]{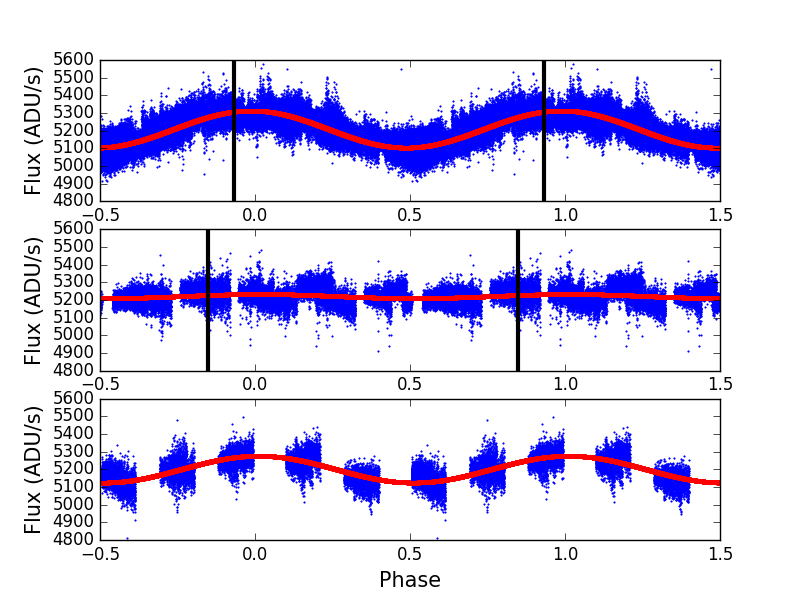}
    \caption{Phase folded data for the three regions of flux variation. Top is the initial active portion, middle is the quiet region and bottom is the following increase in activity. Overlaid in red is the sinusoidal fit for a 59 hour period. The black lines indicate the location of the small flare (top section) and the large flare (middle section).} 
    \label{fig:phase_fold}
\end{figure}

The amplitude of the observed  
spin modulation
evolves with time. We split the lightcurve into three regions of activity, corresponding to an initial active portion, a secondary quiet portion and a final region where the amplitude increases once more. The 59 hour period phase folded data for these regions can be seen in Fig.\,\ref{fig:phase_fold}.
\begin{figure}
	\includegraphics[width=\columnwidth]{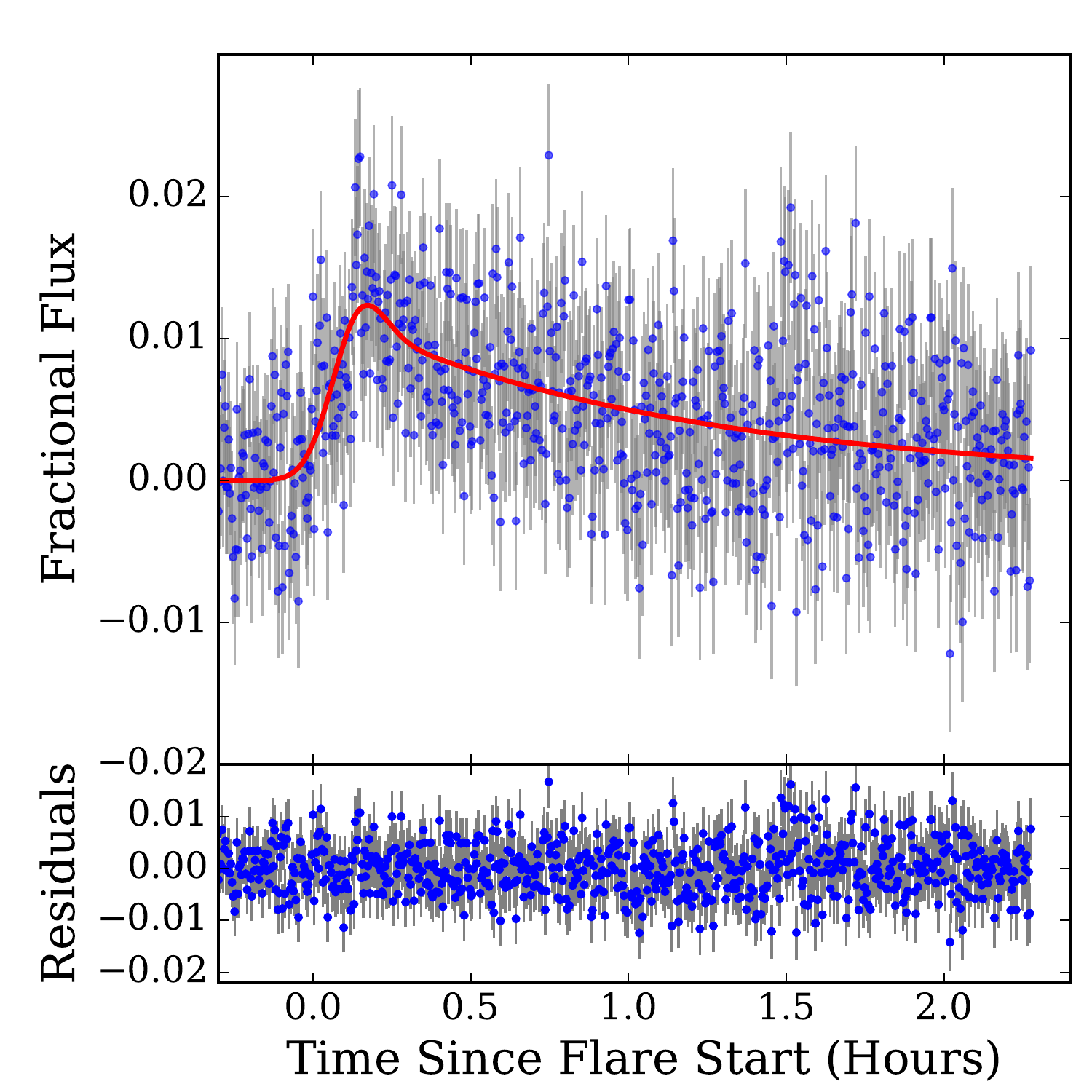}
    \caption{A lower amplitude flare, from 2015 December 17th. The best fitting model is overlaid in red. We note the appearance of substructure at the flare peak and at 1.5 hours and discuss this in Sect.\,\ref{sec:discussion}. The flare start time is given here by where the fit goes above 1\,$\sigma$ above the quiescent flux, as discussed in Sect.\,\ref{sec:flare_amp}}
    \label{fig:secondary_flare}
\end{figure}
\begin{figure}
	\includegraphics[width=\columnwidth]{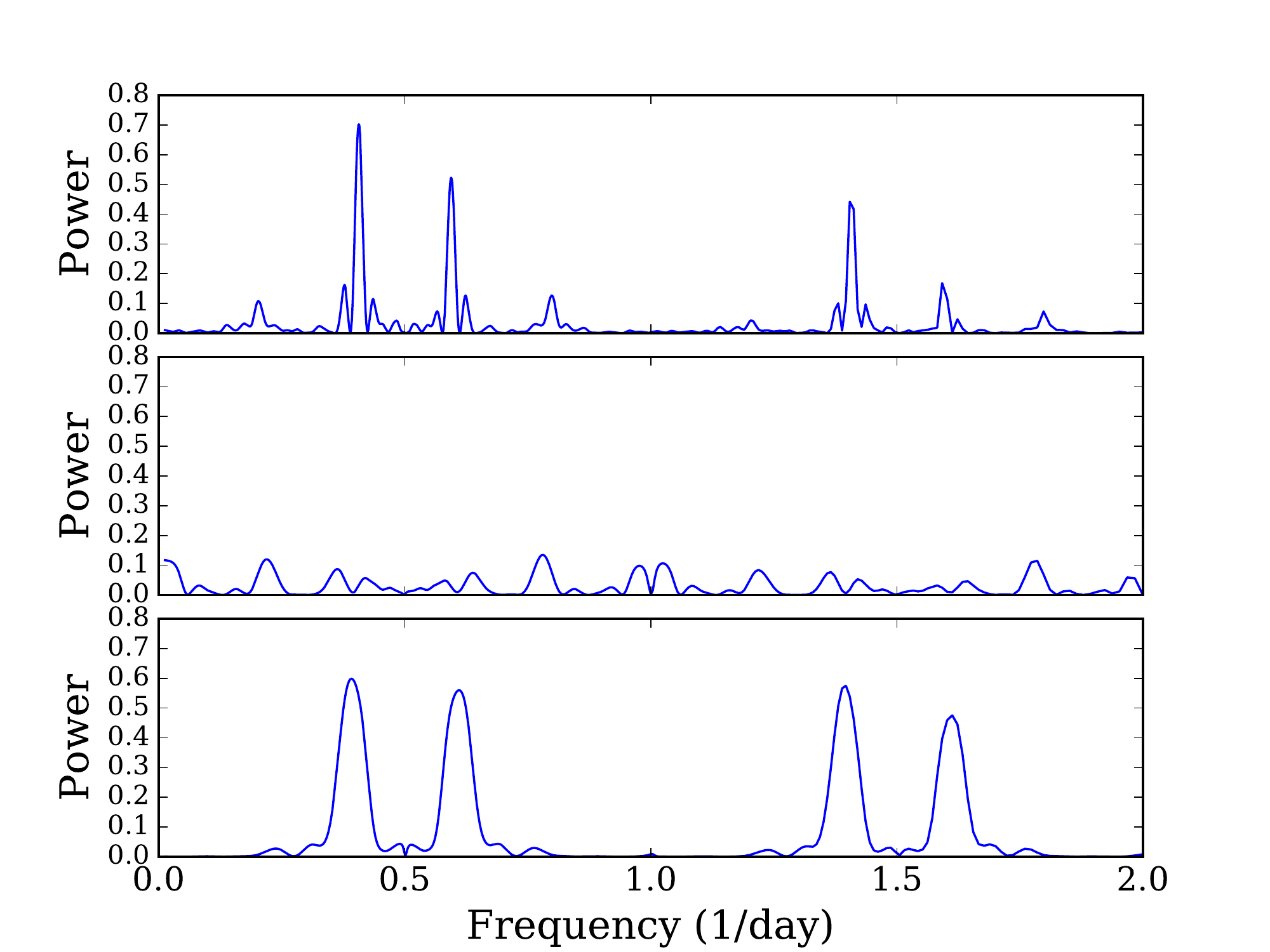}
    \caption{Lomb-Scargle periodograms for each section shown in Fig.\,\ref{fig:phase_fold}. Note the lack of a period detection in the quiescent region.}
	\label{fig:all_ls_plots}
\end{figure}
These regions are plotted in phase relative to the beginning of the lightcurve. The third region has a similar but slightly offset phase from the first, 
as well as less complete phase coverage due to a shorter duration. The durations of each region in the lightcurve are 40, 23 and 17 days respectively. The change in phase, along with the changing flux variation, can be explained by the decay of the original set of star spots and the formation of new ones. Starspot lifetimes have been studied by \citet{Bradshaw14} and for main sequence stars are on month timescales. One example is CoRoT-2, which has a starspot evolution timescale of 31$\pm$15 days \citep[][]{Silva_Valio11}. Consequently we attribute our changing lightcurve modulation to starspot evolution.

We also searched for periodic signals separately in the three light curve regions. The three Lomb-Scargle periodograms are presented in Fig.\,\ref{fig:all_ls_plots}. 
In the second, quiet, period of the light curve we see no evidence for periodic modulation.  
Calculating the modulation period of the third region gives a significantly longer rotation period of 60.87$\pm$0.04 hours (0.39 $d^{-1}$). This offset period suggests that the star  exhibits differential rotation and that the new set of starspots are formed at a different latitude to the original active region. 

We also check where each flare occurs in phase, to search for any relation to the location of the active region. The smaller flare occurs towards the end of the first region, close to maximum optical brightness, 
while the larger flare occurs in the second, quiet region at similar rotation phase (see Fig.\,\ref{fig:phase_fold}). For the smaller flare, this is opposite in phase to the dominant active region. 
Neither flare seems to be located at a rotation phase where a large star spot group is obviously visible, 
and we discuss this further in Sect.\,\ref{sec:discussion}.

\subsection{X-ray Activity} \label{sec:x_ray_result}
As noted in Sect.\,\ref{sec:stellar_prop}, \objname\ has been detected in X-rays with \rosat. 
The detection was made during the \rosat\ all sky survey, and we adopt count rates and hardness ratios from the 2RXS catalog \citep[][]{Boller16}. The \rosat\ PSPC count rate was $0.042\pm0.018\,\rm s^{-1}$ and the hardness ratios in the standard \rosat\ bands were \mbox{HR1=$1.000 \pm 0.325$} and \mbox{HR2=$-0.428 \pm 0.243$}. The HR1 value indicates that the source was detected only in the \rosat\ hard X-ray band (0.5--2.0\,keV) and not in the soft band (0.1--0.4\,keV). 

The \rosat\ PSPC count rate of \objname\ corresponds to a 0.1--2.4\,keV energy flux of $5.7 \times 10^{-13} \textrm{ erg s}^{-1} \textrm{cm}^{-2}$, using energy flux conversion factors determined for coronal sources by \citet{Fleming95}. This flux conversion uses the HR1 hardness ratio to  account of the characteristic temperature of corona, and it has been applied to large samples of stars from the \rosat\ all sky survey by \citet{Schmitt95} and \citet{Huensch}. 

This X-ray flux corresponds to a 0.1--2.4\,keV X-ray luminosity of $L_{X} =1.7 \times 10^{30} \textrm{ erg s}^{-1}$ assuming a distance to \objname\ of 156\,pc that we estimate using
the apparent V magnitude and the expected absolute V magnitude for a G8V star \citep{Gray09}. 
Using the values for \teff\ and $R_{*}$ from Sect.\,\ref{sec:stellar_prop} we find the bolometric luminosity of the star to be $L_{Bol}=1.8 \times 10^{33} \textrm{ erg s}^{-1}$ and hence $\log{L_{X}/L_{Bol}} = -3.1$, which corresponds to saturated X-ray emission \citep{Pizzolato03,Wright11}. 

\begin{figure*}
	\includegraphics[width=\textwidth]{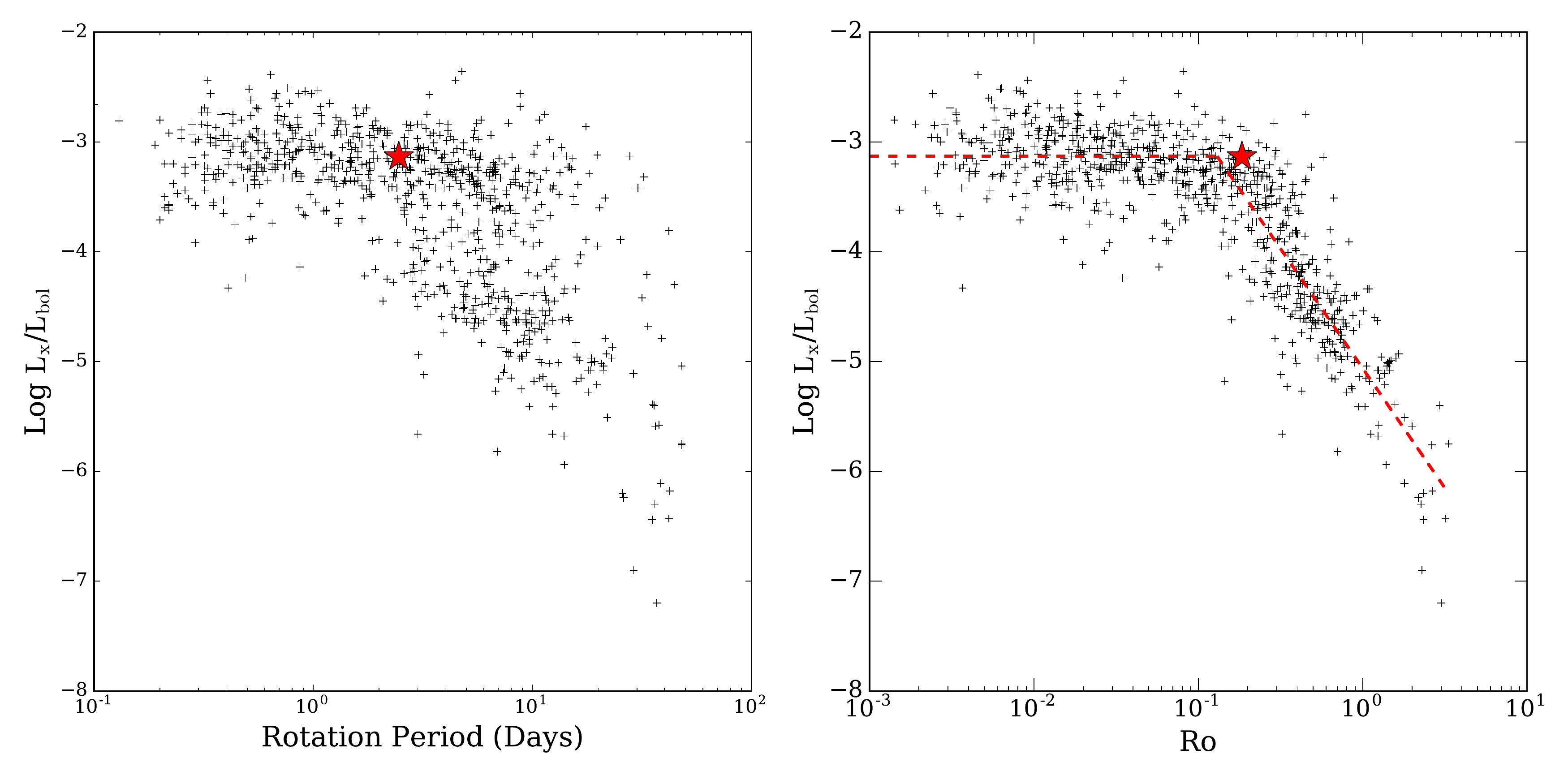}
    \caption{Left: Stellar X-ray to bolometric luminosity ratio vs rotation period for \objname\ with the data from \citet{Wright11}. Right: Same, but for Rossby number. We have also overlaid the power law fit from \citet{Wright11}, with $\beta$=-2.18. \objname\ is shown here with a red star.}
    \label{fig:wright_fig}
\end{figure*}

Combining this X-ray luminosity with our measurement of the stellar rotation period (Sect.\,\ref{sec:stellar_rotation}), we can place \objname\ on the rotation-activity relation of \citet{Wright11}. This is shown in Fig.\,\ref{fig:wright_fig} where \objname\ can be seen to reside close to the break point between saturated X-ray emission and the power law decrease in activity to slower rotation. The Rossby number of 0.18 was calculated using our rotation period and the relation for convective turnover time from \citet{Wright11}. 

Using the relation between the X-ray surface flux and average coronal temperature from \citet{Johnstone15} we estimate an average coronal temperature of 10\,MK. This is similar to the coronal temperature of 7.5\,MK predicted from the rotation period using the relation by \citet{Telleschi05}. 

The lack of detection of \objname\ in the \rosat\ soft band, as well as its relatively large distance, suggests that it may be subject to stronger interstellar absorption than the sample of stars used to determine the flux conversion factors of \citet{Fleming95}. We therefore
double checked our flux estimation using {\sc webpimms}.\footnote{https://heasarc.gsfc.nasa.gov/cgi-bin/Tools/w3pimms/w3pimms.pl} We assumed a characteristic coronal temperature of 7.5\,MK and an interstellar column density equal to the total Galactic columnn the direction of \objname, which is $N_{\rm H}=2\times10^{20}\,\rm cm^{-2}$ \citep{Dickey90,Kalberla05}. The measured \rosat\ PSPC count rate then corresponds to an unabsorbed 0.1--2.4\,keV energy flux of $4.9\times10^{-13}\,\rm erg\,cm^{-2}\,s^{-1}$, which is within 20 percent of our calculation using the flux conversion factors of \citet{Fleming95}. 

\subsection{Flare Modelling} \label{sec:modelling}
We model our flares following a similar method to \citet{Gryciuk17}, who fitted solar flares in soft X-rays. For both flares, we use the convolution of a Gaussian with a double exponential. A Gaussian is used to account for the heating in the flare rise, which has been found to be appropriate for solar flares 
\citep[e.g.][]{Aschwanden98}. A double exponential is used for the decay, accounting for thermal and non-thermal cooling processes, which has been used previously for the decay of stellar flares \citep[e.g.][]{Davenport2014}. A convolution of these Gaussian and exponential functions is then analogous to the heating and cooling processes occurring during the flare \citep[][]{Gryciuk17}.
With this physically motivated model we can utilise the high cadence of \ngts, in particular the flare rise which in the past has been fit using a polynomial or disregarded due to a lack of data points \citep[e.g.][]{Davenport2014,Pugh16}.

Before performing fitting, we inspected the full lightcurve and noted that several nights exhibited behaviour consistent with atmospheric extinction. We identified this trend by using the full lightcurve to fit for a first order atmospheric extinction term. This trend was then removed from the lightcurve, including the nights showing our flares. The nights before and after each flare were used to check the quality of this fit and were found to have the atmospheric extinction successfully removed.
We also account for the flux modulation effects from starspots. To do this we use the preceding two and subsequent two nights and fit a sinusoid at the calculated 59 hour stellar rotation period. With this sinusoid we are able to remove any gradient due to rotation from the night. 
This is required most for the smaller flare, which sits in the first, more active region of the lightcurve (Sect.\,\ref{sec:stellar_rotation}). 

For both flares we perform fitting using an MCMC analysis with the python package \textsc{emcee} \citep{emcee13}, using 500 walkers for 2000 steps and discarding the first 500 as a burn-in. During modelling we have increased our error bars to account for scintillation using the modified Young's approximation with the empirical coefficient for Paranal \citep{Young67,Osborn15}. The best fitting models for the two flares are overlaid on Figs.\,\ref{fig:example_figure}\,\&\,\ref{fig:secondary_flare}. The best fitting parameters presented in Tab.\,\ref{tab:flare_params}. 
\subsubsection{Flare Amplitude and Duration} \label{sec:flare_amp}
To determine the amplitude of each flare we use the maximum value of our fit. For the larger flare this gives a fractional amplitude of 6.9 per cent. For the smaller flare, using the value from the fit gives a fractional amplitude of 1.2 per cent. Inspecting Fig.\,\ref{fig:secondary_flare}, there appears to be impulsive substructure at the flare peak, which is not accounted for in our model. Taking the average of the five data points around the peak gives an peak amplitude of 2.0$\pm$0.3 per cent. 

To obtain a measure of the full duration of the flare, we again make use of our fit. We define the start and end of the flare as the points where 
the model rises and then falls more than 1$\sigma$ above the background flux level, as in \citet{Gryciuk17}. Here $\sigma$ is determined from the quiescent flux before the flare. From this, we determine the flare duration of the larger flare to be 55 minutes. Due to the decreased amplitude to error ratio of the smaller flare, we do not calculate the full flare duration using this method. 
However, we also calculate the flare duration with two additional methods - using its e-folding timescale \citep[as performed in][]{Shibayama13} and its scale time (the duration where the flare is above half the maximum flux value). 
Again, we use our fit for these. For the large and small flare, we calculate the e-folding timescale as 16 and 55.5 minutes respectively, and the scale time as 11 and 42 minutes respectively. With our fit we can also calculate the timescale of the flare rise, using the time from the flare start to the peak of the model. Using this, we calculate the flare rise time as 2.5 minutes for the larger flare. If we use the 1\,$\sigma$ start limit for the smaller flare, we 
estimate
the flare rise as at least 7.4 minutes.
\begin{table}
	\centering
	\begin{tabular}{|l|c|c|}
    \hline
    Property & Large & Small \tabularnewline \hline
   	Energy (erg) & \largeenergy\ & \smallenergy\ \tabularnewline
    Fit Amplitude (per cent) & 6.9 & 1.2 \tabularnewline
    Full Duration (min) & 55 & N/A\tabularnewline
	e-folding duration (min) & 16 & 55.5\tabularnewline
    Scale time (min) & 11 & 42\tabularnewline
    Flare rise (min) & 2.5 & >7.4\tabularnewline
	\hline
	\end{tabular}
    \caption{\label{tab:flare_params}
    Properties of each superflare detected from \objname.}
\end{table}

\subsection{Flare Energy}
The method used to calculate the flare energy is based on that described by \citet{Shibayama13}, and makes the assumption that the flare and star act as blackbody radiators, with the flare having a blackbody spectrum of temperature 9000 $\pm$ 500 K in order to estimate the flare luminosity.
Using the stellar effective temperature and radius from 
Sect.\,\ref{sec:stellar_prop}, we calculate the bolometric energy of the larger and smaller flare to be \largeenergy\ erg and \smallenergy\ erg respectively. It is striking that the smaller flare is only a factor two less energetic despite having an amplitude around six times lower. Comparing to the Carrington event energy of $\approx$ $10^{32}$ ergs \citep{Carrington_Energy}, we can see that each flare had a bolometric energy several hundred times greater than this.

From a total of 422 hours of observation for this star, we have detected two flares. We can use this measurement to estimate the flaring rate for flares above \smallenergy\ erg as approximately 40 per year.

\section{Discussion} \label{sec:discussion}
\subsection{Flare Properties}
We have detected two superflares from the G star \objname\ with high-cadence NGTS optical photometry. 
These
are the first ground-based CCD detections of superflares from a G star. 
Our NGTS observations have much higher cadence than the \kepler\ flare detections, 
allowing us to resolve the flare rise and substructure. 

The larger flare is shown in Fig.\,\ref{fig:example_figure} and was calculated to have a bolometric energy of \largeenergy erg and a fractional amplitude of 6.9 per cent. Due to the increased time resolution of our measurements compared to almost all previous superflare detections, we have been able to fit this flare with a physically-motivated model that includes a Gaussian pulse  to describe the impulsive flare rise (as employed previously for solar flares).
For the decay, our data require two exponential components.
Separate impulsive and gradual decay components been seen previously in some stellar flares,  and 
attributed to decay of blackbody-like emission 
and chromospheric emission respectively \citep[][]{Hawley14,Kowalski_spec13}. We can also see 
that this flare displays a flattening around the peak, or a ``roll-over''. Similar flare peak behaviour has been seen by \citet{Kowalski11} from ULTRACAM observations of the dM3.5e star EQ Peg A. This behaviour is captured in the fitted model as a result of the observed combination of Gaussian heating and exponential cooling. Further, we can identify 
smaller peaks in the decay of the flare, located at approximately 0.7 and 1.0 hour after the flare start in Fig.\, \ref{fig:example_figure}. Structure, or ``bumps'', such as this have been previously identified in flare decays with \kepler\ \citep[e.g][]{Balona15}.

Our model has also been used to fit 
the smaller flare of \objname, shown in Fig.\,\ref{fig:secondary_flare}. 
This flare has a much lower relative amplitude of just 1.2 per cent, 
making it the lowest-amplitude G star flare to have been detected from the ground. 
Despite its low amplitude, this smaller flare has a much slower rise and longer duration than the larger flare (by factors of 3--4) so that it has a high total energy of \smallenergy erg, which is only a factor 2 lower than the larger flare.
When fitting this smaller flare it became apparent that there was additional structure at the flare peak. This can be seen in the residuals of Fig.\,\ref{fig:secondary_flare}, as a small spike lasting approximately one minute. This is a sign of an additional heating pulse at the end of the initial flare rise. In this flare we also detect subtructure around 1.5 hours after the flare start (visible in the residuals). 
The amplitude of the peak at this time is approximately 1 per cent, which is comparable with the amplitude of the main flare. Considering the timing of this substructure relative to the main flare peak, it is likely an example of sympathetic flaring \citep[e.g.][]{Moon02}. 

One advantage of our flare model, combining a Gaussian heating pulse with exponential cooling, 
is that it avoids an arbitary discontinuity between the end of the rise and the beginning of the decay. 
This has generally not been the case with previous stellar flare models, which have tended to include an instantaneous transition between functions describing the rise and decay  \citep[e.g.][]{Davenport2014}. Our model also provides a well-defined measure of the rise timescale, 
allowing for studies of how the flare rise time changes between flares. In this case we see the lower amplitude flare rising much more slowly than the high amplitude example. 
This highlights how wide-field high-cadence surveys such as \ngts\ can  
contribute to the quantitative characterisation of stellar flares.

\subsection{Starspots and Flare Phases} \label{sec:starspots}
Our analysis of the \ngts\ light curve of \objname\ revealed a 59.09$\pm$0.01\,h periodic modulation that we interpret the changing visibility of starspots on the stellar rotation period (Sect.\,\ref{sec:stellar_rotation}). The initial set of starspots appear to decay during the observations, and no spin modulation is detected for an interval of around 23\,d. 
Periodic modulation begins again towards the end of the NGTS observations, and at a slightly longer period, suggesting that the star exhibits differential rotation and that new starspots have emerged at a different latitude. 

Checking where the flares occur in rotation phase reveals that the smallest flare occurs in antiphase to the dominant starspot group, while the larger flare occurs during the quiescent interval of the lightcurve (at a similar spin phase to the first flare). These flare timings are perhaps surprising, as we might expect to see superflares when large active regions are present and visible. 
Instead, our results suggest that the observed superflares do not emerge from the dominant active regions on the stellar surface. Such behaviour is not unprecedented, as observations of the M dwarfs AD~Leo and GJ\,1243 
showed no correlation between stellar flare occurrence and rotational phase \citep{HuntWalker12,Hawley14}. A similar result was found 
for the K dwarf KIC\,5110407, with all but the two strongest flares 
showing no correlation with the most active regions \citep{Roettenbacher13}. 
In these cases it was suggested that the dominant active region might be located at the pole, such that it is always in view and flares can be seen at any spin phase.  
An alternative is that the majority of flares originate from smaller spot groups that do not cause the dominant
flux modulation.
\subsection{Comparison with \kepler} \label{sec:kepler_compare}
\begin{figure}
	\includegraphics[width=\columnwidth]{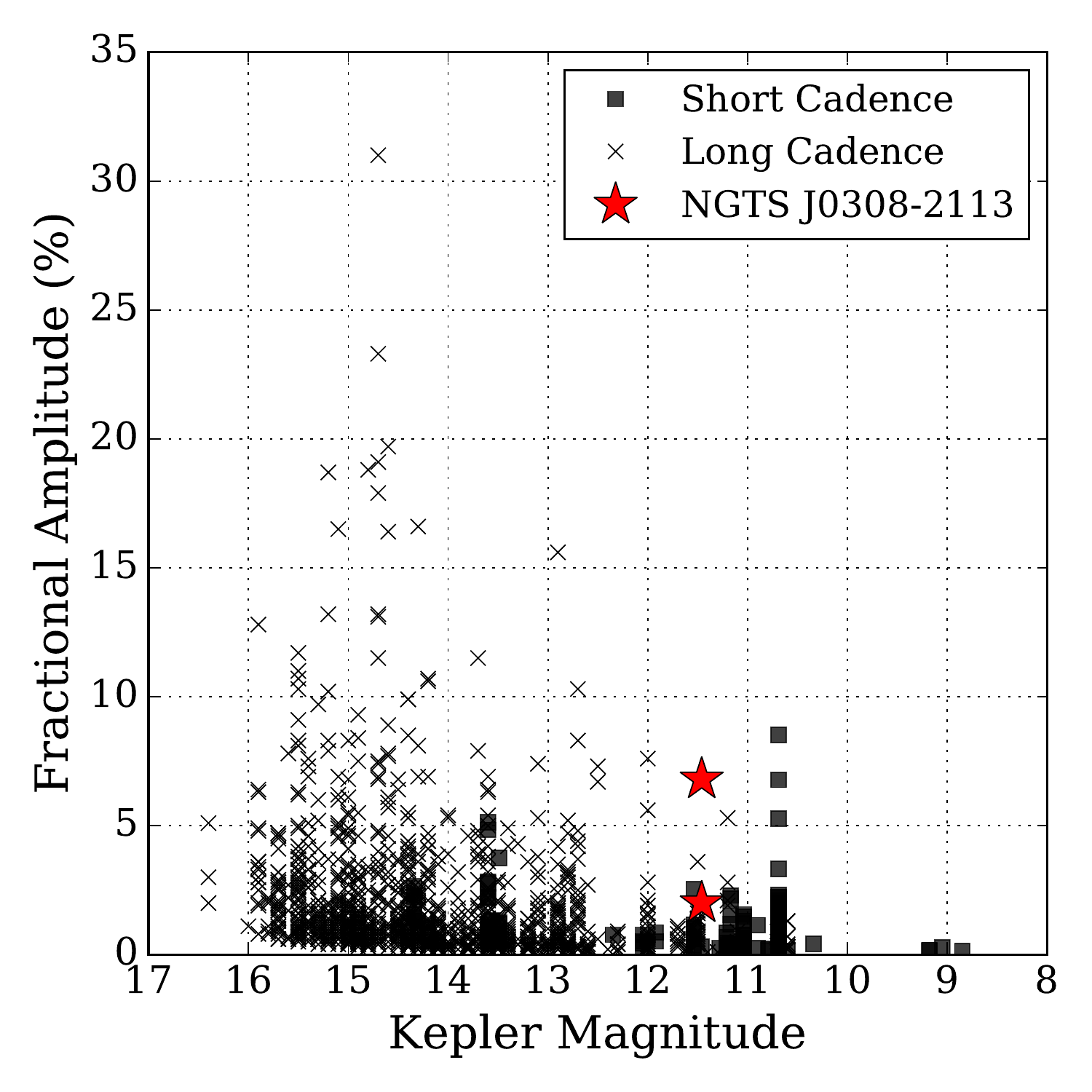}
    \caption{Comparison of the fractional amplitudes and \kepler\ magnitude of our flares from \objname\ (red stars) with G star superflares detected in short cadence (black squares) and long cadence (black crosses) \kepler\ data. 
    When comparing fractional amplitudes between NGTS and \kepler\, it should be noted that the \kepler\ bandpass extends blueward of \ngts.}
    \label{fig:FRAC_MAG}
\end{figure}

In Fig.\,\ref{fig:FRAC_MAG} we compare the superflares of \objname\ with G star superflares detected with \kepler. We use the samples from \citet{Shibayama13} and \citet{Maehara15} for the long and short cadence \kepler\ data respectively. \objname\ has a \kepler\ magnitude of 11.4, calculated using the stellar \textit{g'} and \textit{r'} magnitudes and Eqn.\,2a from \citet{Brown11}. This magnitude makes it one of the brightest G stars seen to exhibit a superflare (see Fig.\,\ref{fig:FRAC_MAG}). 
The larger flare from \objname\ also has a greater amplitude than all but one of those detected in short cadence \kepler\ data. 
This flare also has a shorter duration than most detected with \kepler.
This comparison demonstrates that \ngts\ has a sufficiently wide-field of view and high photometric precision to detect rare and interesting stellar flares from bright stars. Each flare is also observed with higher cadence than has previously been possible. 

\subsection{X-ray Activity}
Thanks to the relative proximity and hence brightness of \objname\ we were able to measure its X-ray luminosity using archival \rosat\ data (Sect.\,\ref{sec:x_ray_result}). 
We found that the star is in the saturated X-ray regime, with $\log L_X / L_{Bol} = -3.1$, and that its X-ray emission is consistent with the rotation activity relation of \citet{Wright11}. It was not detected in the \rosat\ soft X-ray band, likely due to interstellar absorption. Using our measured spin period and the relation for convective turnover time by \citet{Wright11}
we estimated the Rossby number 
of \objname\
to be 0.18. 
Interestingly, this places the star close to the peak of superflare occurrence rates found by 
\citet{Candelaresi14}. 

We can also compare the X-ray luminosity of \objname\ with that of other G stars exhibiting superflares. \citet{Yabuki17} found nine stars with X-ray detections from  
the \kepler\ superflare sample of \citet{Shibayama13}. 
Using these nine X-ray detections they identified a correlation between the largest white-light flare energies (estimated from \kepler\ data) and quiescent $L_{X}$ with
\begin{equation} \label{eq:energy_lx}
\begin{aligned}
    E_{Bol} \propto L_{X}^{1.2\pm^{0.3}_{0.4}}.
\end{aligned}
\end{equation}
Based on this relation, we would expect \objname\ to exhibit flares of energies up to approximately $8 \times 10^{35}$ erg. This is around 13 times greater than the energy of our larger flare, suggesting \objname\ sometimes exhibits even more energetic flares than the examples we have detected with \ngts.
\subsubsection{Maximum Flare Energy}
An alternative method to estimate the potential maximum flare energy is to use the starspot activity. This is done using equation (1) from \citet{Shibata13},
\begin{equation} \label{eq:spot_energy}
\begin{aligned}
    E_{flare} \approx 7\times10^{32}(\textrm{erg})\bigg(\frac{f}{0.1}\bigg)\bigg(\frac{B}{10^{3}\textrm{G}}\bigg)^{2}\bigg(\frac{A_{\textrm{spot}}}{3\times10^{19}\textrm{cm}^{2}}\bigg)^{3/2}
\end{aligned}
\end{equation}
where $f$, $B$ and $A_{\textrm{spot}}$ are the fraction of magnetic energy that can be released as flare energy and the magnetic field strength and area of the starspot respectively. We estimate the starspot area from the lightcurve modulation normalised by the average brightness, following the method of \citet{Notsu13}. We use the region of greatest brightness variation to estimate the area, obtaining a value equivalent to 0.04 of the visible stellar surface. We assume $f$=0.1 \citep{Aschwanden14} and $B$=1000 - 3000G \citep[typical comparison values for solar-type stars e.g.][]{Solanki03, Maehara15} and calculate $E_{flare} = 0.9-8.5\times10^{35}$erg. This estimated value is the same order of magnitude as that calculated from the X-ray data, predicting a flare of greater energy than our largest event.

\subsection{Implications For Exoplanet Habitability}
Understanding the properties of superflares from G stars is important when considering the habitability of Earth-like exoplanets, including those expected to be detected with \PLATO\ \citep{PLATO14}.  
Stellar flares are known to be associated with intense ultraviolet radiation \citep[e.g.][]{Stelzer06, Tsang12}, which 
can reduce levels of 
atmospheric ozone \citep[e.g.][]{Lingam17}
and damage the DNA 
of biological organisms \citep[e.g.][]{Castenholz12}. 
Associated X-ray and extreme-ultraviolet radiation can also erode the planetary atmosphere and drive water loss. 
Stellar flares are also 
associated with
Coronal Mass Ejections (CMEs), 
and while
planetary magnetospheres may protect against the quiescent stellar wind, CMEs can act to compress the magnetosphere and expose the planetary atmosphere to further erosion and dessication \citep[e.g.][]{Kay16,Lammer07}. 

The detections of superflares presented in this paper demonstrate that wide-field ground-based surveys such as \ngts\ are capable of characterising the rates and energies of superflares from G-type stars, despite their relatively low fractional amplitude. Since flare detections with ground-based telescopes can be made and announced in real time it may also be possible to trigger immediate follow up of superflares with larger narrow-field telescopes 
while the flares are still in progress. This has not been possible to date due to the unpredictable nature of superflares and inevitable delays in downlinking and processing data from space telescopes such as \kepler. Real-time follow up of \ngts\ flares might then provide the multi-wavelength observations needed to assess the impact of superflares on potentially habitable exoplanet atmospheres. 

\section{Conclusions}
In this work we have presented the detection of two superflares from the G8 star \longobjname\ using \ngts. These are the first G star superflares detected from the ground using a CCD, and they are among the highest cadence measurements of any superflares to date. We  fit both flares with a model that incorporates a Gaussian heating pulse, 
as seen previously in solar flares, and exponential decay on two timescales. The model fit provides the amplitude, energy and duration of each flare, and we find the two flares have similar total energies despite their different amplitudes and durations. The larger flare has an unusually high amplitude and short duration for a G star superflare. 
Our model also allows us to measure the timescale of the flare rise, an interval that has been 
undersampled 
in previous studies, and we find the longer duration flare has a slower rise. We have also detected substructure in both flares. 

The stellar rotation period of \objname\ was measured to be 59 hours, and we found evidence for differential rotation. The X-ray luminosity of the star was calculated to be $1.7 \times 10^{30} \textrm{ erg s}^{-1}$, with  $\log{L_{X}/L_{Bol}}=-3.1$ implying saturated X-ray emission, as expected for a G8 star with such a short spin period. The Rossby number of 0.18 places \objname\ close to the peak of the occurence rate distribution implied by previous flare detections.  

Our results highlight the potential for wide-field ground-based surveys such as \ngts\ to determine the rates, energies and morphologies of superflares from G stars, despite the modest white-light amplitudes of such flares. Further detections and real-time multi-wavelength follow up will be important in assessing the habitability of Earth-like exoplanets around G stars, including those to be found with \PLATO. 

\section*{Acknowledgements}

This research is based on data collected under the \textit{NGTS} project at the ESO La Silla Paranal Observatory. The NGTS facility is funded by a consortium of institutes consisting of 
the University of Warwick,
the University of Leicester,
Queen's University Belfast,
the University of Geneva,
the Deutsches Zentrum f\" ur Luft- und Raumfahrt e.V. (DLR; under the `Gro\ss investition GI-NGTS'),
the University of Cambridge, together with the UK Science and Technology Facilities Council (STFC; project reference ST/M001962/1). 
JAGJ is supported by an STFC studentship.
PJW, DJA and RGW are supported by STFC consolidated grant ST/P000495/1.
AMB acknowledges the support of the Institute of Advanced Study, University of Warwick and is also supported by STFC consolidated grant ST/P000320/1.
JSJ acknowledges support by Fondecyt grant 1161218 and partial support by CATA-Basal (PB06, CONICYT).
MNG is supported by STFC award reference 1490409 as well as the Isaac Newton Studentship.
CEP acknowledges support from the European
Research Council under the SeismoSun Research Project No. 321141.
The research leading to these results has received funding from the European Research Council under the European Union's Seventh Framework Programme (FP/2007--2013) / ERC Grant Agreement n. 320964 (WDTracer). 




\bibliographystyle{mnras}
\bibliography{references} 





\bsp	
\label{lastpage}
\end{document}